\documentclass[10pt,twocolumn, nofootinbib, aps,prb]{revtex4-1}
\usepackage{amssymb,amsmath}
\usepackage{graphicx}
\usepackage{color}
\usepackage{hyperref}
\usepackage{listings}
\usepackage{natbib}
\usepackage{xfrac}
\usepackage{upgreek}

\begin{document}

\title{\Large Broadband near-infrared astronomical spectrometer calibration and on-sky validation with an electro-optic laser frequency comb}
\author{\textbf{E. Obrzud$^{1,2}$, M. Rainer$^3$, A. Harutyunyan$^4$, B. Chazelas$^2$, M. Cecconi$^4$, A. Ghedina$^4$, E. Molinari$^{4,5}$, S. Kundermann$^1$, S. Lecomte$^{1}$, F. Pepe$^2$, F. Wildi$^2$, F. Bouchy$^2$, T. Herr$^{1,*}$}}
\vspace{2cm}
\affiliation{\vspace{0.5cm} \mbox{$^{1}$Swiss Center for Electronics and Microtechnology (CSEM), Rue de l'Observatoire 58, 2000 Neuch\^atel, Switzerland} \\\mbox{$^{2}$Geneva Observatory, University of Geneva, Chemin des Maillettes 51, 12901 Versoix, Switzerland} \\\mbox{$^{3}$National Institute of Astrophysics (INAF), Astronomical Observatory of Brera, Via Brera 28, 20121 Milano, Italy} \\\mbox{$^{4}$Fundaci\'on Galileo Galilei - INAF, Rambla Jos\'e Ana Fern\'andez P\'erez 7, 38712 Bre\~na Baja, Santa Cruz de Tenerife, Spain} \\\mbox{$^{5}$ INAF - Osservatorio Astronomico di Cagliari, Via della Scienza 5 - 09047 Selargius (CA), Italy} \\ \vspace{0.5cm} $^*$tobias.herr@csem.ch}

\begin{abstract}
The quest for extrasolar planets and their characterisation as well as studies of fundamental physics on cosmological scales rely on capabilities of high-resolution astronomical spectroscopy. A central requirement is a precise wavelength calibration of astronomical spectrographs allowing for extraction of subtle wavelength shifts from the spectra of stars and quasars. Here, we present an all-fibre, 400 nm wide near-infrared frequency comb based on electro-optic modulation with 14.5 GHz comb line spacing.  Tests on the high-resolution, near-infrared spectrometer GIANO-B show a photon-noise limited calibration precision of  $<10$~$\mathrm{\sfrac{{cm}}{s}}$ as required for Earth-like planet detection. Moreover, the presented comb provides detailed insight into particularities of the spectrograph such as detector inhomogeneities and differential spectrograph drifts. The system is validated in on-sky observations of a radial velocity standard star (HD221354) and telluric atmospheric absorption features. The advantages of the system include simplicity, robustness and turn-key operation, features that are valuable at the observation sites.
\end{abstract}

\maketitle

\begin{figure*}[t]
	\centering
	\includegraphics[width = 1\textwidth]{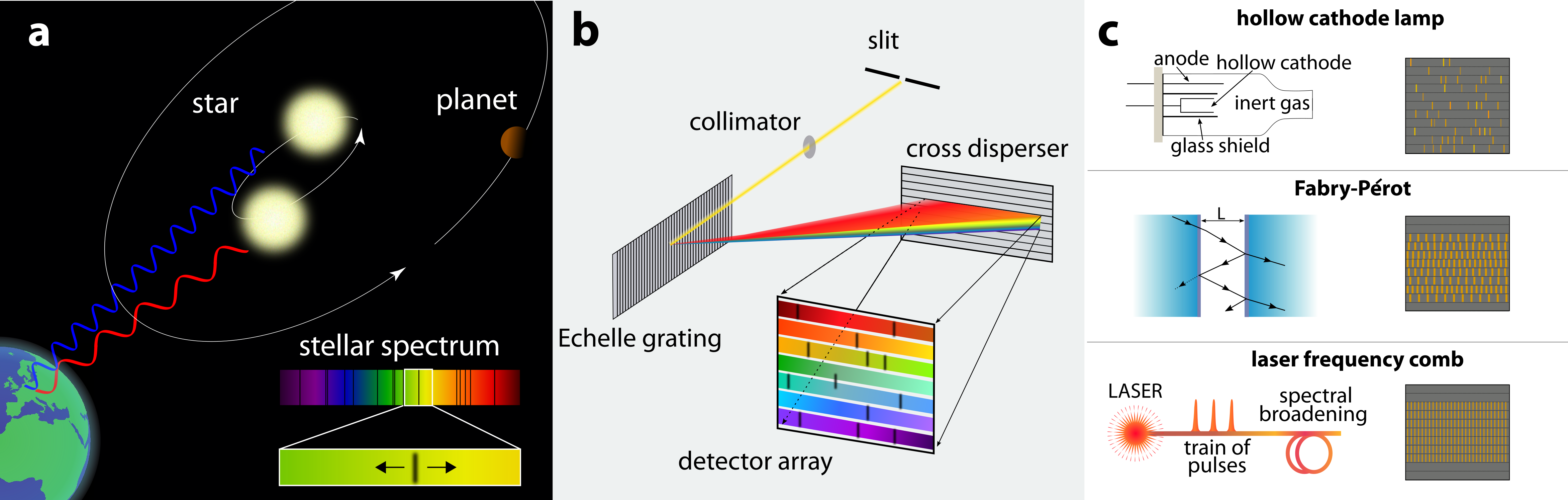}
	\caption{\small \textbf{Radial velocity measurement and astronomical spectrograph calibration.} \textbf{a)} Exoplanet detection with radial velocity method. Due to the gravitational pull of a companion, the host star follows an orbit around the system's centre of mass resulting in periodic Doppler shifts of spectral features. \textbf{b)} Scheme of a cross-dispersed Echelle spectrograph. \textbf{c)} Examples of wavelength calibrators and their spectra: hollow cathode lamps, Fabry-P\'erot cavity and laser frequency combs.}
	\label{fig1}
\end{figure*}

High-resolution spectrographs are invaluable tools in modern astrophysics that allow for a broad scope of applications ranging from exoplanetary science \citep{mayor1995} to cosmology and fundamental physics \citep{uzan2011}. The former relies on astronomical spectrographs for exoplanet detection and characterisation with the radial velocity technique, i.e. detection of Doppler-shifted absorption features in stellar spectra (Fig.\ref{fig1}a). Furthermore, high-resolution spectroscopy allows for probing exoplanet atmospheres \citep{charbonneau2002, wyttenbach2015, martins2015}. Similarly, monitoring spectra of distant astronomical objects enables measurements of the physical constants variability and constraining the nature of the dark matter \citep{liske2008, uzan2011, tamura2016}. Both, radial velocity technique and physical constant measurement, require extremely precise and accurate instruments, e.g. detection of an Earth analogue necessitates a radial velocity precision of 9 $\mathrm{\sfrac{cm}{s}}$ over a period of several years, while a measurement of the Hubble constant requires two decades of quasar monitoring with a precision of 2 $\mathrm{\sfrac{cm}{s}}$. In view of the extend of the science cases that high-precision astronomical spectroscopy addresses, there is a great interest in developing extremely stable instruments capable of detecting radial velocity shifts at the $\mathrm{\sfrac{cm}{s}}$ level.

Modern astronomical spectrographs use a cross-dispersion scheme that divides the observed spectra into Echelle orders on a detector array (Fig.\ref{fig1}b). Wavelength calibration provides a pixel-to-wavelength mapping mapping for the detector array and thus serves as a wavelength reference for the measured spectra. In order to meet the precision requirements, it is essential to use a broadband, well-characterised light source providing a grid of accurately and precisely defined optical lines of uniform intensity with line separations well resolved by the spectrograph. Absolute calibration is crucial for long-term radial velocity monitoring, as it provides a possibility of comparing data from different epochs and instruments. Figure \ref{fig1}c shows different calibration sources and their respective Echelle spectra. Traditionally, gas cells and hollow cathode lamps have been used as wavelength calibrators providing a performance limited to a precision of about 1~$\mathrm{\sfrac{m}{s}}$  due to sparse lines with high dynamic range in intensity, line blending and unstable intensity profile. Alternatively, Fabry-P\'erot interferometers (FPI), fed by an intense white light source produce broadband spectra composed of quasi-equidistant lines (Fig.\ref{fig1}c). Enclosed in a vacuum chamber with mechanical and thermal stabilisation, FPIs can provide a one-night stability of 10~$\mathrm{\sfrac{cm}{s}}$ \citep{wildi2012, bauer2015}. However, a long term stability of both, hollow cathode lamps and FPIs, is not guaranteed; ageing of the lamps or mirror coatings as well as limited long term mechanical stability deteriorate their performance.

A solution that can overcome limitations of standard calibration methods is the laser frequency comb (LFC) technology \citep{udem2002, telle1999, jones2000, cundiff2003} (Fig.\ref{fig1}c). The unprecedented precision and accuracy of these systems has already brought revolutionary progress to the radial velocity field \citep{steinmetz2008, li2008, wilken2012, mccracken2017a, brucalassi2016, ycas2012, doerr2012, hou2015, glenday2015, schettino2011, yi2016, murphy2007, quinlan2010, mccracken2017b}. Frequency combs provide spectra composed of narrow, equally spaced emission lines in the frequency domain, with each optical line frequency $f_n$ described by the relation $\mathrm{f_n = n \cdot f_{rep} + f_{off}}$, where $\mathrm{f_{rep}}$ and $\mathrm{f_{off}}$ are two radio frequencies (RF) - the repetition rate and the carrier-envelope offset frequency, respectively. The control over both $\mathrm{f_{rep}}$ and $\mathrm{f_{off}}$ allows linking the LFC to the cesium-based atomic time and frequency definition resulting in absolute calibration.

There are several approaches for frequency comb generation. Most commonly, LFCs are generated by mode-locked laser (MLLs) that periodically emit femtosecond pulses. Stabilization and referencing of the comb lines to an RF-standard proceeds via $f$-$2f$-self-referencing \citep{udem2002, telle1999, jones2000,cundiff2003}. As the native repetition rate, i.e. the comb's line spacing, of MLLs is typically well below 10 GHz, actively stabilised filtering cavities are used to suppress unwanted modes and hence increase the repetition rate to a value resolvable by astronomical spectrographs. Attention must be paid to the suppression of the side modes so as not to introduce shifts in the apparent frequency of the transmitted mode that may lead to systematic errors \citep{chang2010, chang2012, probst2013, probst2014}. 
A distinct way of generating frequency combs is provided by Kerr-nonlinear optical microresonators \citep{delhaye2007, kippenberg2011}. Recent advances in the field were marked by the first demonstrations of microresonator frequency combs used as wavelength calibrators on astronomical spectrographs \citep{obrzud2017, suh2018}.

\begin{figure*}[]
	\centering
	\includegraphics[width=1\textwidth]{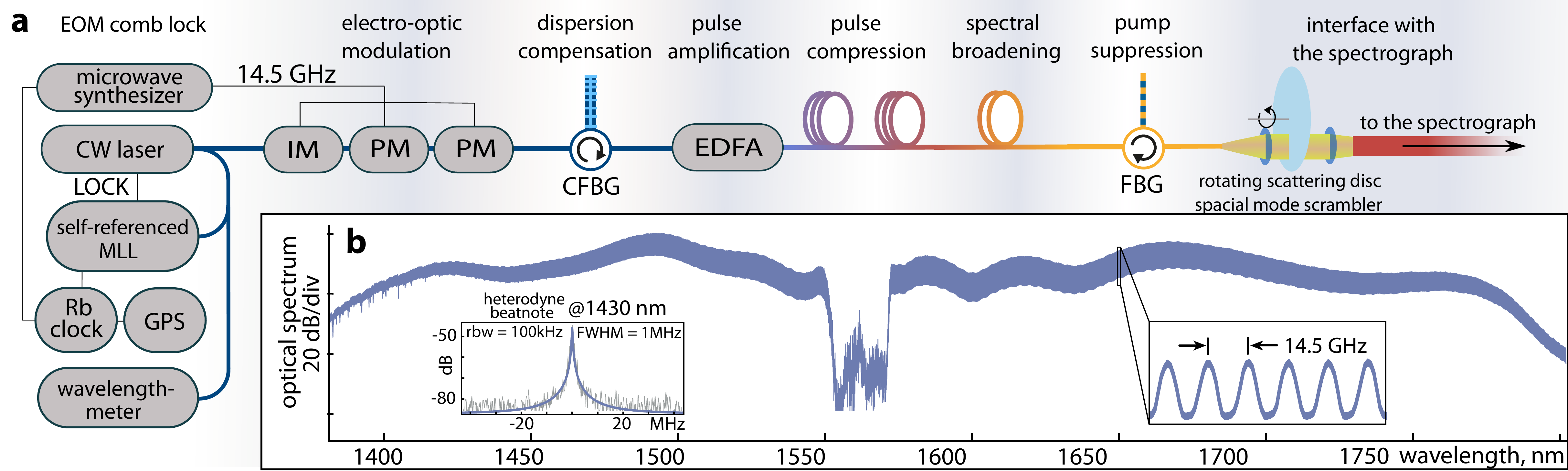}
	\caption{\small \textbf{Electro-optical modulation-based frequency comb set-up.} \textbf{a)} Scheme representing the EOM-based LFC. MLL - mode-locked laser, IM - intensity modulator, PM - phase modulator, CFBG - chirped fibre-Bragg grating, EDFA - erbium-doped fibre amplifier, FBG - fibre-Bragg grating  \textbf{b)} Resulting spectrum spanning $~$400 nm within 20dB with line spacing of 14.5 GHz. Inset: Heterodyne beatnote of the comb with an additional laser measured at 1430 nm showing a width of 1 MHz corresponding to the linewidth of the 1430 nm laser and providing an upper limit to the EOM LFC linewidth.}
	\label{fig2}
\end{figure*}

Electro-optical modulation (EOM) of a continuous wave (CW) laser is another alternative technique for generating LFCs \citep{kobayashi1988, murata2000, morohashi2008, ishizawa2011, beha2017, yi2016, torrescompany2016}. Here, phase modulation of a CW component results in sideband generation with a line spacing given by an external modulation source driving the electro-optic modulators. When driven by a microwave source, EOM-based combs allow for comb line separation in excess of 10 GHz, which is directly compatible with the requirements of astronomical spectrographs. A pioneering demonstration by Yi et al. \citep{yi2016} showed an operation of a electro-optical frequency comb with a repetition rate of 12~GHz on the CSHELL and NIRSPEC spectrographs. More recently, an advanced EOM-based astrocomb system with a line spacing of 30 GHz in the short-wavelength near-infrared was used to demonstrate the intrinsic stability of the HPF spectrograph of $<$10~$\mathrm{\sfrac{cm}{s}}$ over several days. Notably, this demonstration included the telescope optics, dual-fibre spectrograph illumination as well as data reduction \citep{metcalf2018}.

In this article, we present a turn-key EOM-based astrocomb with a line spacing of 14.5 GHz spanning over a wavelength range from 1400 nm to 1800 nm. With this system, we demonstrate a calibration precision reaching $<10$~$\mathrm{\sfrac{cm}{s}}$ as well as validation in on-sky stellar observation. Specifically, the performance of the EOM LFC was tested on the GIANO-B high-resolution spectrograph. In addition to wavelength calibration and spectrograph drift measurements, relative radial velocity measurements of HD221354 were carried out. The LFC is able to precisely track spectrograph drifts, trace subtle instrumental features and provide precise and accurate wavelength calibration for stellar observations.  

\section*{Results}

\subparagraph{Setup.}

The EOM-based laser frequency comb consists of a 1560 nm CW laser sent through a series of synchronised intensity and two phase modulators driven by a microwave (MW) signal generator at 14.5 GHz (Fig.\ref{fig2}a). While phase modulation imprints a chirp on the light wave, the intensity modulator carves out the half period of the phase modulation with a quadratic phase change in time. With this scheme, multiple sidebands are generated around the initial CW laser line resulting in a frequency comb with a flat-top spectrum of 5 nm span. By compensating the dispersion via a chirped fibre Bragg grating (CFBG) (approximately 4 ps/nm), the formation of a train of pulses with a duration of about 2 ps is achieved. These pulses are amplified in an erbium-doped fibre amplifier (EDFA) reaching an average power of 3.5 W. Next, nonlinear optical pulse compression in length-optimized stretches of normal and anomalous dispersion optical fibres results in pulses of 150 fs duration with peak power exceeding 1 kW (2.3 W of average power). Owing to this high peak power, a short-length highly-nonlinear fiber (HNLF) is used for nonlinear spectral broadening generating a coherent broadband comb spectrum from 1400 nm to 1800 nm (Fig.\ref{fig2}b). The HNLF was chosen such that a flat spectral envelope is achieved over the entire spectral range. This permits using the comb without additional filters or waveshapers for broadband equalisation of the spectral wings, which would otherwise significantly increase the system's complexity. The low-intensity feature in the central part of the generated spectrum results from suppression of the pump component by a fibre-Bragg grating (FBG). This feature can be avoided in future work by using an adapted FBG with reduced reflection. With an average output power of 0.8 W, the power per mode is at the level of 300 $\upmu$W, exceeding by several orders of magnitude the photon flux needed for spectrograph calibration. 

\begin{figure*}[!ht]
	\centering
	\includegraphics[width=1\textwidth]{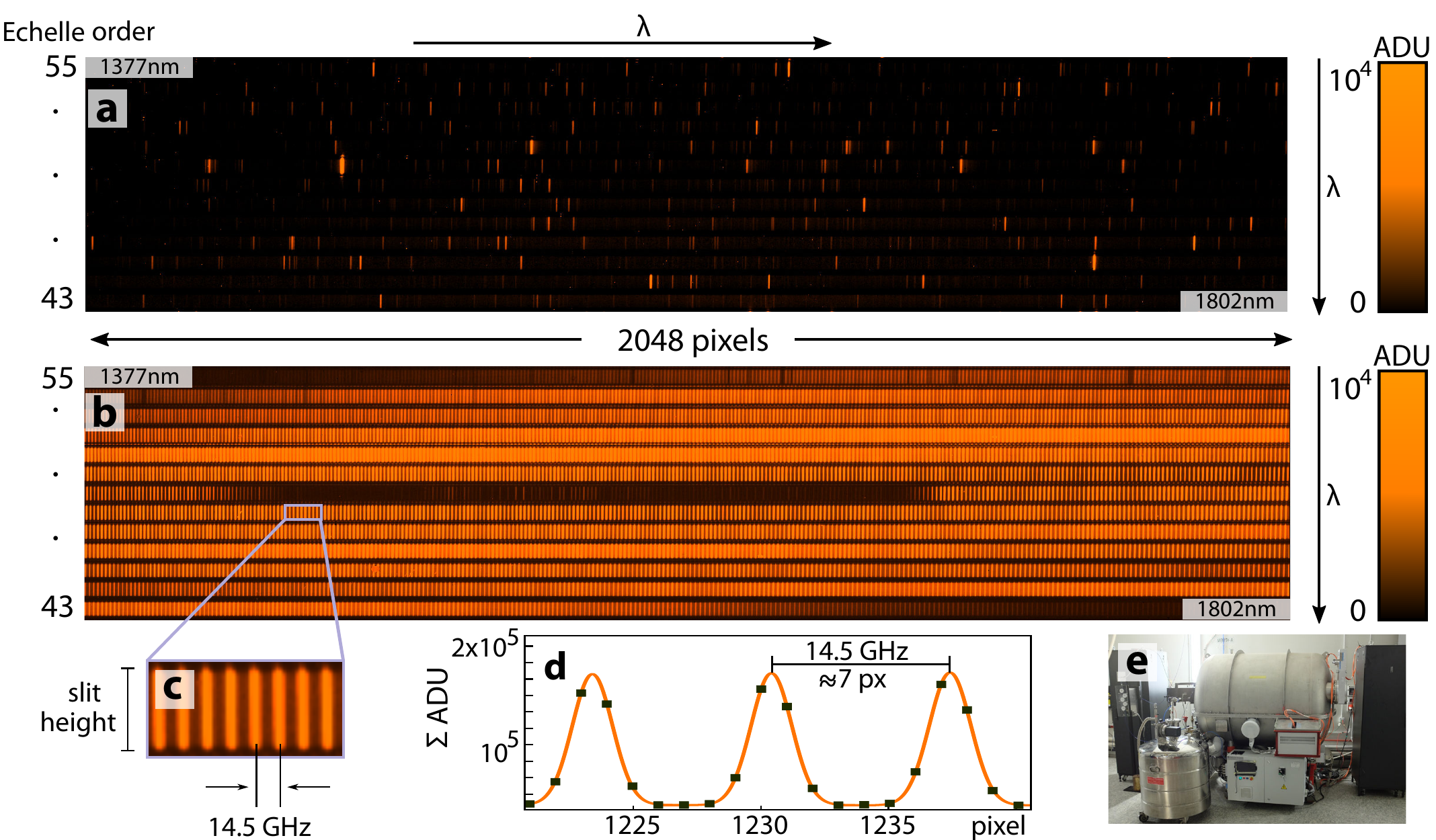}
	\caption{\small \textbf{Spectrometer raw data.} \textbf{a)} Echelle spectrum of a uranium-neon hollow cathode lamp. \textbf{b)} Echelle spectrum of the electro-optic modulation-based laser frequency comb. The dark area in the middle of the 49$^{th}$ order corresponds to the suppressed pump laser. \textbf{c)} The vertical extend of each comb line is given by the slit height, the horizontal profile by the spectrometer's point-spread-function. \textbf{d)} 1-dimensional data obtained by summing the signal in the analogue-to-digital units (ADU) along the slit with a Gaussian fit. \textbf{e)} GIANO-B spectrograph at the Telescopio Nazionale Galileo \citep{claudi2016}.}
	\label{fig3}
\end{figure*}

In order to provide absolute frequency calibration, the CW laser frequency and the MW signal generator are phase-coherently linked to the 10 MHz RF signal of a GPS-disciplined rubidium atomic clock. In order to establish the RF-to-optical-link for the CW laser, a portable home-build self-referenced 100 MHz MLL is used. A wavelength-meter provides an approximate measurement of the CW laser wavelength such that the 100 MHz ambiguity of the MLL can be lifted. The EOM-based frequency comb shows a high line contrast across the entire spectrum as indicated by a heterodyne beatnote with an additional 1430 nm diode laser in the far-out wing of the comb, where the linewidth is expected to be the largest \citep{beha2017}(cf. Fig2b, inset). The beatnote's width of 1 MHz is an upper limit on the LFC linewidth and indeed corresponds to the linewidth of the CW diode laser. Coupling to the spectrograph is achieved by a free-space transition from the comb's output single-mode fibre to the multi-mode fibre guiding the light to the spectrometer. A rotating scattering disc spatial mode scrambler between the single and multi-mode fibres is used to remove modal noise, which would otherwise introduce calibration errors \citep{baudrand2001}.

\begin{figure*}[!ht]
	\centering
	\includegraphics[width=1\textwidth]{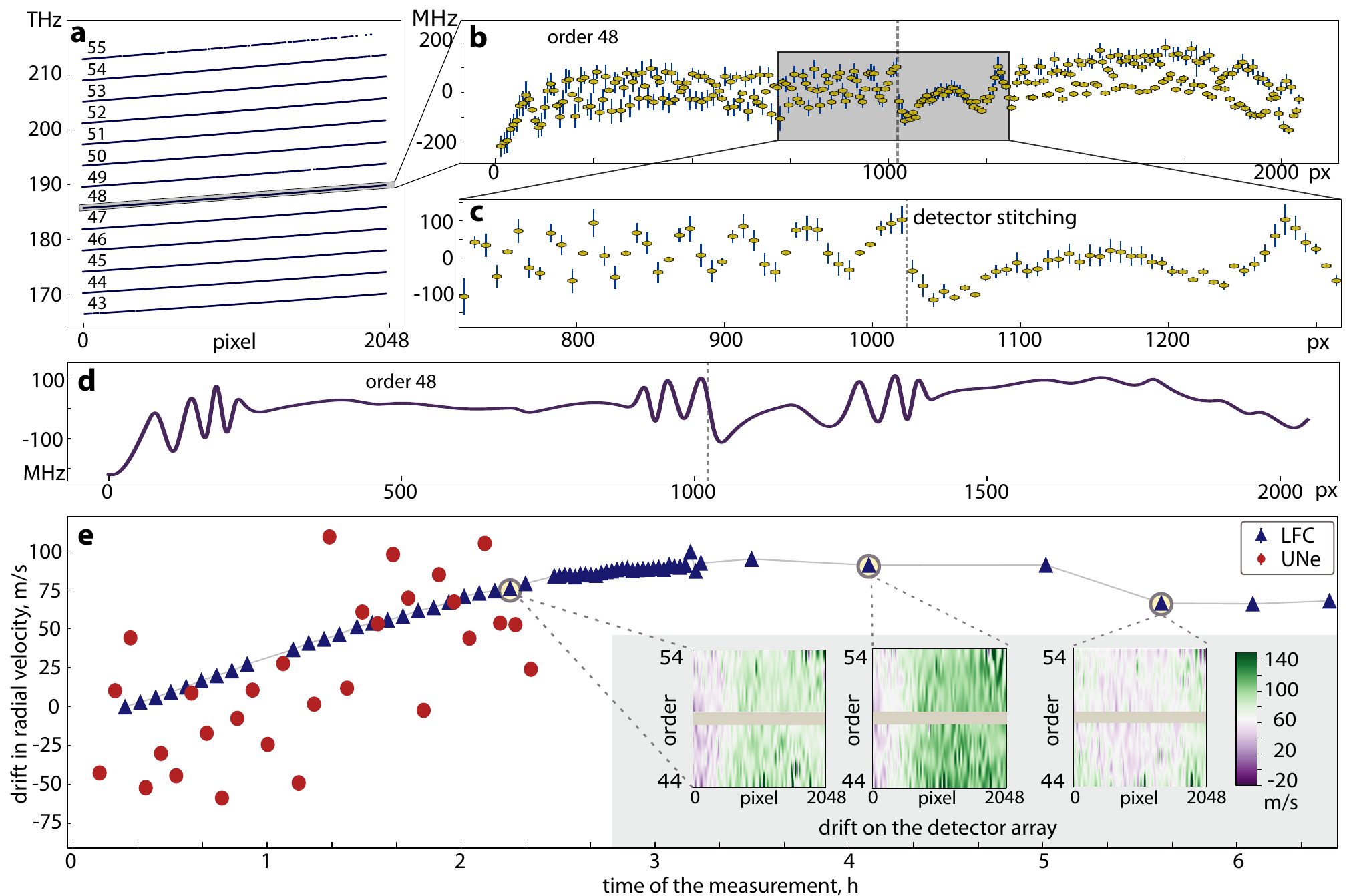}
	\caption{\small \textbf{Wavelength calibration and spectrograph drift measurement.} \textbf{a)} Frequency as a function of pixel for each spectrograph order as derived from assigning comb-frequencies to comb-positions. \textbf{b)} The same as a) but for order 48 with a third order polynomial subtracted for clarity. Dashed grey line indicates detector stitching point. \textbf{c)} Zoom into the inner part of b) showing a discontinuity due to the detector stitching. \textbf{d)} Wavelength calibration for order 48 obtained with the BOK free-knot spline. \textbf{e)} Spectrograph drift measured with the LFC (blue triangles) and the uranium-neon hollow-cathode lamp (red dots). Insets show the colour-coded differential drift observed on the detector array for three LFC exposures. Grey area indicates the order 49 not taken into account for the analysis.}
	\label{fig4}
\end{figure*}

The system is all-fibre-based with polarisation maintaining components that ensure stable operation despite temperature fluctuations or mechanical vibration. It employs highly reliable off-the-shelf fibre-optical components allowing stable operation immediately after system start-up. The prototype system is mounted on three easily transportable 45 cm x 45 cm breadboards. These properties are a great asset from the vantage point of astronomy, where low-complexity systems and low-maintenance operation are indispensable for routine operation at remote sites such as astronomical observatories.

\subparagraph{Wavelength calibration of the GIANO-B spectrograph.}

The demonstration of the EOM-based LFC was performed during several days in November 2017 on the GIANO-B spectrograph (Fig.\ref{fig3}e) mounted on the 3.6 m Telescopio Nazionale Galileo (TNG) at the Roque de los Muchachos Observatory in La Palma, Spain \citep{oliva2012}. GIANO-B is a NIR Echelle spectrograph with a resolution of 50'000 covering the wavelength range from 0.95 $\upmu$m to 2.45 $\upmu$m. Observed spectra are composed of 50 spectral orders organised on a detector array of 2048 x 2048 pixels (Hawaii2RG). The extent of a single optical frequency on the detector is given horizontally by the spectrograph's point-spread-function (PSF) and vertically by the size of the entrance slit.
An example of a uranium-neon (U-Ne) hollow-cathode lamp spectrum, a standard calibration source, is shown in Figure \ref{fig3}a (120 s exposure time). The spectrum exhibits sparse lines of variable intensity, which limit the precision of the wavelength solution. Figure \ref{fig3}b presents the same spectral region for the EOM-based LFC after a 10 s exposure. The LFC provides a dense grid of lines that is well resolved by the spectrograph (Fig.\ref{fig3}c). The observed LFC line-shapes are given by the spectrometer's point-spread-function, which is many orders of magnitude larger than the width of the LFC's modes (cf. inset in Fig.\ref{fig2}b).

Deriving a wavelength calibration starts with extracting 1-dimensional data for each Echelle order by summing the signal (analogue-to-digital units, ADU) over the inner 20 pixels along the vertical direction of the slit (Fig.\ref{fig3}c and d). The position of each comb line is determined by fitting a Gaussian function (corresponding to the instrument's PSF) followed by ascribing an exact optical frequency determined on the basis of the known frequency comb parameters. The uncertainty on the fitting is generally below 100 MHz ($~$2\% of the PSF's linewidth), the error being due to fundamental photon noise \citep{bouchy2001}. The result of assigning comb-frequencies to pixel-positions is shown in Figure \ref{fig4}a. Subtracting a third order polynomial from the data (for visibility) reveals subtle structures related to particularities of the spectrograph optics and detector (Fig.\ref{fig4}b and c). Most notably, one can observe a discontinuity in the middle of each order (Fig.\ref{fig4}c). This results from the GIANO-B detector being actually a mosaic of four 1024 x 1024 pixel detectors. In the middle of every order there is a discontinuity in the frequency-vs-pixel function due to the detector stitching inducing micrometer deviations in the regular pixel arrangement.

\begin{figure*}[]
	\centering
	\includegraphics[width=1\textwidth]{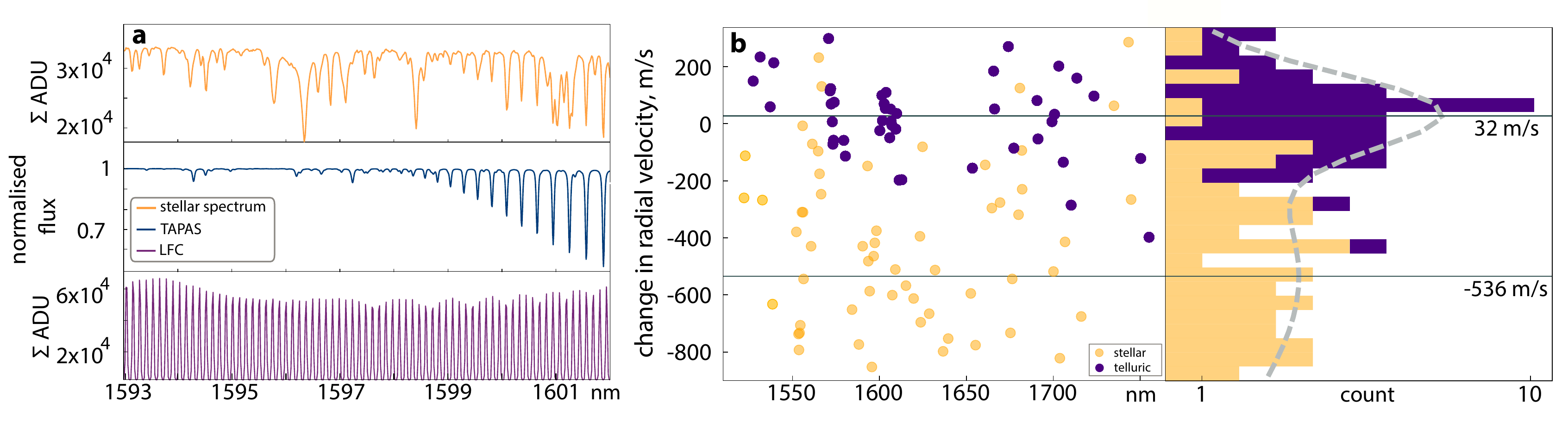}
	\caption{\small \textbf{Relative radial velocities for stellar and telluric lines.} \textbf{a)} Top: Spectrum of HD221354. Middle: Telluric spectrum from TAPAS. Bottom: LFC spectrum. \textbf{b)} Left: Change of apparent radial velocities of HD221354 absorption lines (orange markers) and telluric lines (purple markers) as a function of their wavelength.  Right: Histogram of obtained per-line values for stellar and telluric lines (one outlier at +500~$\mathrm{\sfrac{m}{s}}$ not shown). The dashed line indicates the fit of a bimodal Gaussian distribution yielding apparent relative radial velocity shifts of $+$32$~\mathrm{\sfrac{m}{s}}$ for the telluric and $-$536~$\mathrm{\sfrac{m}{s}}$ for the stellar spectrum. The latter reflects the difference in Earth's barycentric velocity between the two observational epochs of -544~$\mathrm{\sfrac{m}{s}}$.}
	\label{fig7}
\end{figure*}

Finally, the wavelength solution, i.e. a complete pixel-to-optical frequency mapping is derived by fitting the data shown in Figure \ref{fig4}a. We employ bounded optimal knots (BOK) free-knot splines method that is based on the algorithm developed by Molinari et al. \citep{molinari2004}. The Python implementation of this algorithm is provided by the PyCS Python package \citep{tewes2013, bonvin2016}. Regular spline regression divides the data into parts and fits each section using a low order polynomial that is relied to other sections by knots. The advantage of the BOK free-knot-splines is its improved knot-position optimisation. The algorithm minimizes the $\chi^2$ by adjusting spline knots positions and fits sections of data with a third order polynomial, taking also into account errors at each data point. This algorithm proves to be an excellent tool in fitting highly variable data (e.g. in the analysis of quasar time series), capturing well the structures in the data without over-fitting (cf. Fig.\ref{fig4}d). The (BOK) free-knot splines are applied to each order separately and generate a frequency vs pixel function for each of 2048 pixels for every order covered by the LFC. This provides a wavelength solution which corrects for spectrograph imperfections.

One way of determining the calibration precision is comparing two consecutive wavelength calibrations. Here, we consider two wavelength solutions that are separated in time by 2 minutes. For each order, a mean value of the pixelwise differences between the wavelength calibrations is calculated; the standard error for each order is estimated based on the number of statistically independent calibration points (i.e. the number of comb lines). The global drift of the spectrograph is given by a weighted mean over all orders and the precision by the standard error of the weighted mean. The measured spectrometer drift was 70 $\pm$~14~$\mathrm{\sfrac{cm}{s}}$ indicating a global precision of $<10$~$\mathrm{\sfrac{cm}{s}}$ for an individual wavelength solution, fulfilling in principle the requirements for Earth-like planet detection (if all pixels are treated as independent values a precision of 3.5~$\mathrm{\sfrac{cm}{s}}$ would result). This value is an upper-limit estimation on the wavelength calibration precision, as it also contains the intrinsic spectrograph instability. We point out that reaching this level of precision in practice largely depends on the observation conditions, the coupling of the star light into the spectrometer as well as the global and differential drift rate of the spectrograph. Note that order 49 is excluded from the analysis due to the spectral hole stemming from the not-optimized pump suppression.

\subparagraph{GIANO-B drift measurements.}

Next, the spectrograph's drift is investigated by taking frequent LFC exposures during several hours. The first part of the experiment consisted in alternating the LFC and U-Ne exposures followed by a series of only LFC exposures. The results of the drift measurement are presented in Fig.\ref{fig4}e. Both, the LFC and U-Ne lamp based wavelength calibrations agree with each other and show the same drift with the LFC resulting in significantly more precise values (reduced scatter).

Additionally, we performed an analysis of a possible wavelength-dependent differential drift of the spectrograph. The insets in Fig.\ref{fig4}e show a detector view for three chosen LFC exposures well separated in time. The results show that the right part of the detector drifts at a different rate than the left part during a phase of global linear drift. The wavelength-dependent drift pattern changed after the spectrograph steadied and started to drift in the other direction. The observed effect may be due mechanical drift of the spectrograph and/or thermal regulation of the detector. These results show that the LFC is not only capable of following very precisely the spectrograph's global drift, but it also provides a detailed map of differential drifts which are a valuable insight into the characteristics of the spectrograph.

\subparagraph{On-sky observations.}

We could benefit from two slots for observations of HD221354 -  a nearby K-type star with a visible magnitude of $~$6.7 and a constant radial velocity within $\sim$2~$\mathrm{\sfrac{m}{s}}$ \citep{howard2011, bouchy2013}. The observations were separated in time by 48 hours and each stellar observation (10~min exposure) was bracketed by two LFC exposures (10~s exposure each). A linear drift of the instrument during the stellar exposure is assumed, so that the final wavelength calibration for the stellar data is the mean value of the respective bracketing exposures. Deriving radial velocities from the stellar data is not a straightforward task, as the observed spectra are composed of not only stellar absorption features but also telluric lines - absorption features caused by the Earth's atmosphere. Separating telluric and stellar absorption features is a first step to undertake for extraction of radial velocities. To do so, we use the synthetic telluric spectra as provided by TAPAS (Transmissions Atmosph\'eriques Personnalis\'ees Pour l'AStronomie)\citep{bertaux2012, bertaux2014} the specific time, location and meteorological conditions. Figure \ref{fig7}a shows a section of the stellar, TAPAS and LFC sprectra.The TAPAS spectrum accurately predicts positions and contrast of telluric lines making it a well-suited tool for generating a mask with which the observed spectrum can be divided into stellar and telluric parts. Next, for both interference-free stellar and telluric spectral portions, the algorithm finds positions of lines and fits a Gaussian. Relative radial velocities between the two observations are determined for every absorption line; for the telluric radial velocity distribution sigma clipping is performed to reject the outliers. Figure \ref{fig7}b shows the distribution of the measured radial velocities for both stellar and telluric lines and the stacked histogram for the radial velocity measurements. Both stellar and telluric radial velocity distributions are joined and a bimodal distribution is fitted (right panel of \ref{fig7}b). The final results quoted are the medians of the output parameters of the bimodal distribution fit, for which multiple realisations were preformed by marginalizing over the initial fit parameters and number of histogram bins to ensure robustness.

Based on the observed lines, an apparent radial velocity shift of -536~$\mathrm{\sfrac{m}{s}}$ is measured for HD221354, while 32~$\mathrm{\sfrac{m}{s}}$ is found for the telluric lines. These are close to the expected values for a correctly calibrated spectrometer. The apparent change in the star's radial velocity results from the difference between the Earth's barycentric velocity, equal to -544~$\mathrm{\sfrac{m}{s}}$ between the two observations, whereas close to zero shift is expected for the telluric lines in the atmosphere (co-moving with the telescope and only subject to changes in meteorological conditions). The error on the radial velocity shifts for the telluric lines results mainly from the error introduced by the spectrograph slit where a few tens of $\mathrm{\sfrac{m}{s}}$ is expected due to inhomogeneous illumination of the slit caused by imperfect telescope pointing and guiding. The distribution of relative radial velocities for stellar lines is significantly larger than for the telluric lines, which we attribute to the blending between stellar lines with imperfectly masked telluric lines. We anticipate that a dedicated mask for stellar lines, that was not available for this star, can significantly reduce the scatter found for the stellar lines to the level obtained for the telluric lines. It is noteworthy that with regard to the on-sky observations the calibration performance is not limited by the LFC but corresponds to the inherent performance limit of the slit-illuminated GIANO-B spectrograph in conjunction with telescope pointing and observational conditions\citep{bouchy2004} as well as challenges in the data analysis.

\section*{Conclusion}
To summarise, we demonstrated broadband astronomical spectrograph calibration of an EOM based laser frequency comb. Notably, the frequency comb is adjustment-free owing to its polarization maintaining all-fibre design that does not include free-space elements such as filtering cavities, spectral equalisers or waveshapers. Turn-key operation enables start-up times of a few minutes from a powered-off state to a fully, phase-coherent GPS- and atomic-clock-referenced operation. The actual calibrated wavelength range of nearly 400 nm results in a photon-noise-limited calibration precision of $<10$~$\mathrm{\sfrac{cm}{s}}$ on the high-resolution near-infrared GIANO-B spectrograph significantly outperforming the currently used standard calibrator (uranium-neon hollow-cathode lamp). In particular, the precise measurement of the spectrograph drift reveals fine structures of the wavelength calibration, i.e. detector inhomogeneities and differential instrument drifts, impossible to grasp with standard calibration methods. Importantly, we could also for the first time validate the performance of an EOM-comb in on-sky observations of the radial velocity standard HD221354 and telluric atmospheric absorption features. 

Without compromising on simplicity and robustness, an even larger spectral span could be covered by using stronger phase modulation and higher-power optical amplification along with all-fibre femtosecond pulse compression, as we demonstrate here. In this case an optical microwave-noise suppression stage would need to be inserted into the system in order to achieve narrow comb lines in the far out wings of the spectrum, known from octave spanning combs \citep{beha2017}. Alternatively, with the existence of CW lasers,  amplifiers and electro-optic modulators at both edges of the NIR band, it is conceivable to duplicate the system at 1 $\upmu$m and/or 2 $\upmu$m in order to increase the spectral coverage. In conclusion, the demonstrated EOM-technology is a uniquely suited tool for spectrometer calibration not only for its performance but notably its robustness and low operational complexity. It is of immediate relevance for the next generation of astronomical precision instruments but also of high interest to a large number of existing instruments that currently use gas lamps as calibrators.

\subparagraph{Acknowledgements}

This work was supported by the Swiss National Science Foundation, the NCCR-PlanetS and NCCR-PlanetS Technology Platform, the Canton of Neuch\^atel and INAF Progetto Premiale WOW. D.F. Phillips and the Havard-Smithsonian Center for Astrophysics' astrocomb team are acknowledged for providing the RF reference signal (GPS-disciplined Rb-clock).


\begin{thebibliography}{10}

\bibitem{mayor1995}
M.~{Mayor} and D.~{Queloz}.
\newblock {A Jupiter-mass companion to a solar-type star}.
\newblock {\em \nat}, 378:355--359, November 1995.

\bibitem{uzan2011}
J.-P. {Uzan}.
\newblock {Varying Constants, Gravitation and Cosmology}.
\newblock {\em Living Reviews in Relativity}, 14:2, March 2011.

\bibitem{charbonneau2002}
D.~{Charbonneau}, T.~M. {Brown}, R.~W. {Noyes}, and R.~L. {Gilliland}.
\newblock {Detection of an Extrasolar Planet Atmosphere}.
\newblock {\em \apj}, 568:377--384, March 2002.

\bibitem{wyttenbach2015}
A.~{Wyttenbach}, D.~{Ehrenreich}, C.~{Lovis}, S.~{Udry}, and F.~{Pepe}.
\newblock {Spectrally resolved detection of sodium in the atmosphere of HD189733b with the HARPS spectrograph}.
\newblock {\em Astronomy \& Astrophysics}, 577:A62, May 2015.

\bibitem{martins2015}
J.~H.~C. {Martins}, N.~C. {Santos}, P.~{Figueira}, J.~P. {Faria},
  M.~{Montalto}, I.~{Boisse}, D.~{Ehrenreich}, C.~{Lovis}, M.~{Mayor},
  C.~{Melo}, F.~{Pepe}, S.~G. {Sousa}, S.~{Udry}, and D.~{Cunha}.
\newblock {Evidence for a spectroscopic direct detection of reflected light
  from 51 Pegasi b}.
\newblock {\em Astronomy \& Astrophysics}, 576:A134, April 2015.

\bibitem{liske2008}
J.~{Liske}, A.~{Grazian}, E.~{Vanzella}, M.~{Dessauges}, M.~{Viel},
  L.~{Pasquini}, M.~{Haehnelt}, S.~{Cristiani}, F.~{Pepe}, G.~{Avila},
  P.~{Bonifacio}, F.~{Bouchy}, H.~{Dekker}, B.~{Delabre}, S.~{D'Odorico},
  V.~{D'Odorico}, S.~{Levshakov}, C.~{Lovis}, M.~{Mayor}, P.~{Molaro},
  L.~{Moscardini}, M.~T. {Murphy}, D.~{Queloz}, P.~{Shaver}, S.~{Udry},
  T.~{Wiklind}, and S.~{Zucker}.
\newblock {Cosmic dynamics in the era of Extremely Large Telescopes}.
\newblock {\em Monthly Notices Of The Royal Astronomical Society}, 386:1192--1218, May 2008.

\bibitem{tamura2016}
N.~{Tamura}, N.~{Takato}, A.~{Shimono}, Y.~{Moritani}, K.~{Yabe},
  Y.~{Ishizuka}, A.~{Ueda}, Y.~{Kamata}, H.~{Aghazarian}, S.~{Arnouts},
  G.~{Barban}, R.~H. {Barkhouser}, R.~C. {Borges}, D.~F. {Braun}, M.~A. {Carr},
  P.-Y. {Chabaud}, Y.-C. {Chang}, H.-Y. {Chen}, M.~{Chiba}, R.~C.~Y. {Chou},
  Y.-H. {Chu}, J.~{Cohen}, R.~P. {de Almeida}, A.~C. {de Oliveira}, L.~S. {de
  Oliveira}, R.~G. {Dekany}, K.~{Dohlen}, J.~B. {dos Santos}, L.~H. {dos
  Santos}, R.~{Ellis}, M.~{Fabricius}, D.~{Ferrand}, D.~{Ferreira},
  M.~{Golebiowski}, J.~E. {Greene}, J.~{Gross}, J.~E. {Gunn}, R.~{Hammond},
  A.~{Harding}, M.~{Hart}, T.~M. {Heckman}, C.~M. {Hirata}, P.~{Ho}, S.~C.
  {Hope}, L.~{Hovland}, S.-F. {Hsu}, Y.-S. {Hu}, P.-J. {Huang}, M.~{Jaquet},
  Y.~{Jing}, J.~{Karr}, M.~{Kimura}, M.~E. {King}, E.~{Komatsu}, V.~{Le Brun},
  O.~{Le F{\`e}vre}, A.~{Le Fur}, D.~{Le Mignant}, H.-H. {Ling}, C.~P.
  {Loomis}, R.~H. {Lupton}, F.~{Madec}, P.~{Mao}, L.~S. {Marrara}, C.~{Mendes
  de Oliveira}, Y.~{Minowa}, C.~{Morantz}, H.~{Murayama}, G.~J. {Murray},
  Y.~{Ohyama}, J.~{Orndorff}, S.~{Pascal}, J.~M. {Pereira}, D.~{Reiley},
  M.~{Reinecke}, A.~{Ritter}, M.~{Roberts}, M.~A. {Schwochert}, M.~D.
  {Seiffert}, S.~A. {Smee}, L.~{Sodre}, D.~N. {Spergel}, A.~J. {Steinkraus},
  M.~A. {Strauss}, C.~{Surace}, Y.~{Suto}, N.~{Suzuki}, J.~{Swinbank}, P.~J.
  {Tait}, M.~{Takada}, T.~{Tamura}, Y.~{Tanaka}, L.~{Tresse}, O.~{Verducci},
  D.~{Vibert}, C.~{Vidal}, S.-Y. {Wang}, C.-Y. {Wen}, C.-H. {Yan}, and
  N.~{Yasuda}.
\newblock {Prime Focus Spectrograph (PFS) for the Subaru telescope: overview,
  recent progress, and future perspectives}.
\newblock In {Ground-based and Airborne Instrumentation for Astronomy VI},
  volume 9908 of {SPIE proceedings}, page 99081M, August 2016.

\bibitem{wildi2012}
F.~{Wildi}, B.~{Chazelas}, and F.~{Pepe}.
\newblock {A passive cost-effective solution for the high accuracy wavelength
  calibration of radial velocity spectrographs}.
\newblock In {Ground-based and Airborne Instrumentation for Astronomy IV},
  volume 8446 of {SPIE proceedings}, page 84468E, September 2012.

\bibitem{bauer2015}
F.~F. {Bauer}, M.~{Zechmeister}, and A.~{Reiners}.
\newblock {Calibrating echelle spectrographs with Fabry-P{\'e}rot etalons}.
\newblock {\em Astronomy \& Astrophysics}, 581:A117, September 2015.

\bibitem{udem2002}
T.~{Udem}, R.~{Holzwarth}, and T.~W. {H{\"a}nsch}.
\newblock {Optical frequency metrology}.
\newblock {\em \nat}, 416:233--237, March 2002.

\bibitem{telle1999}
H.~R. {Telle}, G.~{Steinmeyer}, A.~E. {Dunlop}, J.~{Stenger}, D.~H. {Sutter},
  and U.~{Keller}.
\newblock {Carrier-envelope offset phase control: A novel concept for absolute
  optical frequency measurement and ultrashort pulse generation}.
\newblock {\em Applied Physics B: Lasers and Optics}, 69:327--332, October
  1999.

\bibitem{jones2000}
D.~J. {Jones}, S.~A. {Diddams}, J.~K. {Ranka}, A.~{Stentz}, R.~S. {Windeler},
  J.~L. {Hall}, and S.~T. {Cundiff}.
\newblock {Carrier-Envelope Phase Control of Femtosecond Mode-Locked Lasers and
  Direct Optical Frequency Synthesis}.
\newblock {\em Science}, 288:635--640, April 2000.

\bibitem{cundiff2003}
Steven~T. Cundiff and Jun Ye.
\newblock Colloquium: Femtosecond optical frequency combs.
\newblock {\em Rev. Mod. Phys.}, 75:325--342, Mar 2003.

\bibitem{steinmetz2008}
T.~{Steinmetz}, T.~{Wilken}, C.~{Araujo-Hauck}, R.~{Holzwarth}, T.~W.
  {H{\"a}nsch}, L.~{Pasquini}, A.~{Manescau}, S.~{D'Odorico}, M.~T. {Murphy},
  T.~{Kentischer}, W.~{Schmidt}, and T.~{Udem}.
\newblock {Laser Frequency Combs for Astronomical Observations}.
\newblock {\em Science}, 321:1335, September 2008.

\bibitem{li2008}
C.-H. {Li}, A.~J. {Benedick}, P.~{Fendel}, A.~G. {Glenday}, F.~X.
  {K{\"a}rtner}, D.~F. {Phillips}, D.~{Sasselov}, A.~{Szentgyorgyi}, and R.~L.
  {Walsworth}.
\newblock {A laser frequency comb that enables radial velocity measurements
  with a precision of 1cms$^{-1}$}.
\newblock {\em \nat}, 452:610--612, April 2008.

\bibitem{wilken2012}
T.~{Wilken}, G.~L. {Curto}, R.~A. {Probst}, T.~{Steinmetz}, A.~{Manescau},
  L.~{Pasquini}, J.~I. {Gonz{\'a}lez Hern{\'a}ndez}, R.~{Rebolo}, T.~W.
  {H{\"a}nsch}, T.~{Udem}, and R.~{Holzwarth}.
\newblock {A spectrograph for exoplanet observations calibrated at the
  centimetre-per-second level}.
\newblock {\em \nat}, 485:611--614, May 2012.

\bibitem{mccracken2017a}
R.~A. {McCracken}, {\'E}.~{Depagne}, R.~B. {Kuhn}, N.~{Erasmus}, L.~A.
  {Crause}, and D.~T. {Reid}.
\newblock {Wavelength calibration of a high resolution spectrograph with a
  partially stabilized 15-GHz astrocomb from 550 to 890 nm}.
\newblock {\em Optics Express}, 25:6450, March 2017.

\bibitem{brucalassi2016}
A.~{Brucalassi}, F.~{Grupp}, H.~{Kellermann}, L.~{Wang}, F.~{Lang-Bardl},
  N.~{Baisert}, S.~M. {Hu}, U.~{Hopp}, and R.~{Bender}.
\newblock {Stability of the FOCES spectrograph using an astro-frequency comb as
  calibrator}.
\newblock In {\em Ground-based and Airborne Instrumentation for Astronomy VI},
  volume 9908 of {\em SPIE}, page 99085W, August 2016.

\bibitem{ycas2012}
G.~G. {Ycas}, F.~{Quinlan}, S.~A. {Diddams}, S.~{Osterman}, S.~{Mahadevan},
  S.~{Redman}, R.~{Terrien}, L.~{Ramsey}, C.~F. {Bender}, B.~{Botzer}, and
  S.~{Sigurdsson}.
\newblock {Demonstration of on-sky calibration of astronomical spectra using a
  25 GHz near-IR laser frequency comb}.
\newblock {\em Optics Express}, 20:6631, March 2012.

\bibitem{doerr2012}
H.-P. {Doerr}, T.~{Steinmetz}, R.~{Holzwarth}, T.~{Kentischer}, and
  W.~{Schmidt}.
\newblock {A Laser Frequency Comb System for Absolute Calibration of the VTT
  Echelle Spectrograph}.
\newblock {\em Solar Physics}, 280:663--670, October 2012.

\bibitem{hou2015}
Lei Hou, Hai-Nian Han, Wei Wang, Long Zhang, Li-Hui Pang, De-Hua Li, and Zhi-Yi
  Wei.
\newblock A 23.75-ghz frequency comb with two low-finesse filtering cavities in
  series for high resolution spectroscopy.
\newblock {\em Chinese Physics B}, 24(2):024213, 2015.

\bibitem{glenday2015}
Alexander~G. Glenday, Chih-Hao Li, Nicholas Langellier, Guoqing Chang, Li-Jin
  Chen, Gabor Furesz, Alexander~A. Zibrov, Franz K\"{a}rtner, David~F.
  Phillips, Dimitar Sasselov, Andrew Szentgyorgyi, and Ronald~L. Walsworth.
\newblock Operation of a broadband visible-wavelength astro-comb with a
  high-resolution astrophysical spectrograph.
\newblock {\em Optica}, 2(3):250--254, Mar 2015.

\bibitem{schettino2011}
G.~{Schettino}, E.~{Oliva}, M.~{Inguscio}, C.~{Baffa}, E.~{Giani}, A.~{Tozzi},
  and P.~C. {Pastor}.
\newblock {Optical Frequency Comb as a general-purpose and wide-band
  calibration source for astronomical high resolution infrared spectrographs}.
\newblock {\em Experimental Astronomy}, 31:69--81, August 2011.

\bibitem{yi2016}
X.~{Yi}, K.~{Vahala}, J.~{Li}, S.~{Diddams}, G.~{Ycas}, P.~{Plavchan},
  S.~{Leifer}, J.~{Sandhu}, G.~{Vasisht}, P.~{Chen}, P.~{Gao}, J.~{Gagne},
  E.~{Furlan}, M.~{Bottom}, E.~C. {Martin}, M.~P. {Fitzgerald}, G.~{Doppmann},
  and C.~{Beichman}.
\newblock {Demonstration of a near-IR line-referenced electro-optical laser
  frequency comb for precision radial velocity measurements in astronomy}.
\newblock {\em Nature Communications}, 7:10436, January 2016.

\bibitem{murphy2007}
M.~T. {Murphy}, T.~{Udem}, R.~{Holzwarth}, A.~{Sizmann}, L.~{Pasquini},
  C.~{Araujo-Hauck}, H.~{Dekker}, S.~{D'Odorico}, M.~{Fischer}, T.~W.
  {H{\"a}nsch}, and A.~{Manescau}.
\newblock {High-precision wavelength calibration of astronomical spectrographs
  with laser frequency combs}.
\newblock {\em Monthly Notices Of The Royal Astronomical Society}, 380:839--847, September 2007.

\bibitem{quinlan2010}
F.~{Quinlan}, G.~{Ycas}, S.~{Osterman}, and S.~A. {Diddams}.
\newblock {A 12.5 GHz-spaced optical frequency comb spanning 400 nm for
  near-infrared astronomical spectrograph calibration}.
\newblock {\em Review of Scientific Instruments}, 81(6):063105--063105, June
  2010.

\bibitem{mccracken2017b}
R.~A. {McCracken}, J.~M. {Charsley}, and D.~T. {Reid}.
\newblock {Decade of astrocombs: recent advances in frequency combs for
  astronomy}.
\newblock {\em Optics Express}, 25:15058, June 2017.


\bibitem{chang2010}
G.~{Chang}, C.-H. {Li}, D.~F. {Phillips}, R.~L. {Walsworth}, and F.~X.
  {K{\"a}rtner}.
\newblock {Toward a broadband astro-comb: effects of nonlinear spectral
  broadening in optical fibers}.
\newblock {\em Optics Express}, 18:12736, June 2010.

\bibitem{chang2012}
G.~{Chang}, C.-H. {Li}, D.~F. {Phillips}, A.~{Szentgyorgyi}, R.~L. {Walsworth},
  and F.~X. {K{\"a}rtner}.
\newblock {Optimization of filtering schemes for broadband astro-combs}.
\newblock {\em Optics Express}, 20:24987, October 2012.

\bibitem{probst2013}
R.~A. {Probst}, T.~{Steinmetz}, T.~{Wilken}, H.~{Hundertmark}, S.~P. {Stark},
  G.~K.~L. {Wong}, P.~S.~J. {Russell}, T.~W. {H{\"a}nsch}, R.~{Holzwarth}, and
  T.~{Udem}.
\newblock {Nonlinear amplification of side-modes in frequency combs}.
\newblock {\em Optics Express}, 21:11670, May 2013.

\bibitem{probst2014}
R.~A. {Probst}, G.~{Lo Curto}, G.~{Avila}, B.~L. {Canto Martins}, J.~R. {de
  Medeiros}, M.~{Esposito}, J.~I. {Gonz{\'a}lez Hern{\'a}ndez}, T.~W.
  {H{\"a}nsch}, R.~{Holzwarth}, F.~{Kerber}, I.~C. {Le{\~a}o}, A.~{Manescau},
  L.~{Pasquini}, R.~{Rebolo-L{\'o}pez}, T.~{Steinmetz}, T.~{Udem}, and Y.~{Wu}.
\newblock {A laser frequency comb featuring sub-cm/s precision for routine
  operation on HARPS}.
\newblock In {\em Ground-based and Airborne Instrumentation for Astronomy V},
  volume 9147 of {\em SPIE proceedings}, page 91471C, July 2014.

\bibitem{delhaye2007}
P.~{Del'Haye}, A.~{Schliesser}, O.~{Arcizet}, T.~{Wilken}, R.~{Holzwarth}, and
  T.~J. {Kippenberg}.
\newblock {Optical frequency comb generation from a monolithic microresonator}.
\newblock {\em \nat}, 450:1214--1217, December 2007.

\bibitem{kippenberg2011}
T.~J. {Kippenberg}, R.~{Holzwarth}, and S.~A. {Diddams}.
\newblock {Microresonator-Based Optical Frequency Combs}.
\newblock {\em Science}, 332:555, April 2011.

\bibitem{obrzud2017}
E.~{Obrzud}, M.~{Rainer}, A.~{Harutyunyan}, M.~H. {Anderson}, M.~{Geiselmann},
  B.~{Chazelas}, S.~{Kundermann}, S.~{Lecomte}, M.~{Cecconi}, A.~{Ghedina},
  E.~{Molinari}, F.~{Pepe}, F.~{Wildi}, F.~{Bouchy}, T.~J. {Kippenberg}, and
  T.~{Herr}.
\newblock {A Microphotonic Astrocomb}.
\newblock {\em ArXiv e-prints}, December 2017.

\bibitem{suh2018}
M.-G. {Suh}, X.~{Yi}, Y.-H. {Lai}, S.~{Leifer}, I.~S. {Grudinin}, G.~{Vasisht},
  E.~C. {Martin}, M.~P. {Fitzgerald}, G.~{Doppmann}, J.~{Wang}, D.~{Mawet},
  S.~B. {Papp}, S.~A. {Diddams}, C.~{Beichman}, and K.~{Vahala}.
\newblock {Searching for Exoplanets Using a Microresonator Astrocomb}.
\newblock {\em ArXiv e-prints}, January 2018.

\bibitem{kobayashi1988}
T.~{Kobayashi}, H.~{Yao}, K.~{Amano}, Y.~{Fukushima}, and A.~{Morimoto}.
\newblock {Optical pulse compression using high-frequency electrooptic phase
  modulation}.
\newblock {\em IEEE Journal of Quantum Electronics}, 24:382--387, February
  1988.

\bibitem{murata2000}
H.~{Murata}, A.~{Morimoto}, T.~{Kobayashi}, and S.~{Yamamoto}.
\newblock {Optical pulse generation by electrooptic-modulation method and its
  application to integrated ultrashort pulse generators}.
\newblock {\em IEEE Journal of Selected Topics in Quantum Electronics},
  6:1325--1331, November 2000.

\bibitem{morohashi2008}
I.~{Morohashi}, T.~{Sakamoto}, H.~{Sotobayashi}, T.~{Kawanishi}, I.~{Hosako},
  and M.~{Tsuchiya}.
\newblock {Widely repetition-tunable 200 fs pulse source using a Mach
  Zehnder-modulator-based flat comb generator and dispersion-flattened
  dispersion-decreasing fiber}.
\newblock {\em Optics Letters}, 33:1192, May 2008.

\bibitem{torrescompany2016}
V.~{Torres-Company}, A.M.~{Weiner}.
\newblock {Optical frequency comb technology for ultra-broadband radio-frequency photonics}.
\newblock {\em Laser \& Photonics Reviews}, 8 (3), 368-393, 2013.

\bibitem{ishizawa2011}
A.~{Ishizawa}, T.~{Nishikawa}, A.~{Mizutori}, H.~{Takara}, H.~{Nakano},
  T.~{Sogawa}, A.~{Takada}, and M.~{Koga}.
\newblock {Generation of 120-fs laser pulses at 1-GHz repetition rate derived
  from continuous wave laser diode}.
\newblock {\em Optics Express}, 19:22402, November 2011.

\bibitem{beha2017}
K.~{Beha}, D.~C. {Cole}, P.~{Del'Haye}, A.~{Coillet}, S.~A. {Diddams}, and
  S.~B. {Papp}.
\newblock {Electronic synthesis of light}.
\newblock {\em Optica}, 4:406--411, 2017.

\bibitem{metcalf2018}
A.~J.~{Metcalf}, C.~{Bender}, S.~{Blakeslee}, W.~{Brand}, D.~{Carlson}, S.~A.~{Diddams}, C.~{Fredrick}, S.~{Halverson}, F.~{Hearty}, D.~{Hickstein}, J.~{Jennings}, S.~{Kanodia}, K.~{Kaplan}, E.~{Lubar}, S.~{Mahadevan}, A.~{Monson}, J.~{Ninan}, C.~{Nitroy}, S.~{Papp}, L.~{Ramsey}, P.~{Robertson}, A.~{Roy}, C.~{Schwab}, K.~{Srinivasan}, G.~K.~{Stefansson}, R.~{Terrien}.
\newblock {Infrared Astronomical Spectroscopy for Radial Velocity Measurements with 10 cm/s Precision}.
\newblock {\em Conference on Lasers and Electro-Optics}, JTh5A.1, May 2018.

\bibitem{baudrand2001}
J.~{Baudrand} and G.~A.~H. {Walker}.
\newblock {Modal Noise in High-Resolution, Fiber-fed Spectra: A Study and
  Simple Cure}.
\newblock {\em Publications of the Astronomical Society of the Pacific}, 113:851--858, July 2001.

\bibitem{claudi2016}
R.~{Claudi}, S.~{Benatti}, I.~{Carleo}, A.~{Ghedina}, G.~{Micela},
  E.~{Molinari}, E.~{Oliva}, A.~{Tozzi}, and {Giarps Team}.
\newblock {GIARPS: the VIS-NIR high precision radial velocity facility TNG}.
\newblock In {\em Frontier Research in Astrophysics II, held 23-28 May, 2016 in
  Mondello (Palermo), Italy (FRAPWS2016).}, page~70, May 2016.

\bibitem{oliva2012}
E.~{Oliva}, L.~{Origlia}, R.~{Maiolino}, C.~{Baffa}, V.~{Biliotti}, P.~{Bruno},
  G.~{Falcini}, V.~{Gavriousev}, F.~{Ghinassi}, E.~{Giani}, M.~{Gonzalez},
  F.~{Leone}, M.~{Lodi}, F.~{Massi}, I.~{Mochi}, P.~{Montegriffo}, M.~{Pedani},
  E.~{Rossetti}, S.~{Scuderi}, M.~{Sozzi}, and A.~{Tozzi}.
\newblock {The GIANO spectrometer: towards its first light at the TNG}.
\newblock In {\em Ground-based and Airborne Instrumentation for Astronomy IV},
  volume 8446 of {\em SPIE proceedings}, page 84463T, September 2012.

\bibitem{bouchy2001}
F.~{Bouchy}, F.~{Pepe}, and D.~{Queloz}.
\newblock {Fundamental photon noise limit to radial velocity measurements}.
\newblock {\em Astronomy \& Astrophysics}, 374:733--739, August 2001.

\bibitem{molinari2004}
Nicolas Molinari, Jean{-}Fran{\c{c}}ois Durand, and Robert Sabatier.
\newblock Bounded optimal knots for regression splines.
\newblock {\em Computational Statistics {\&} Data Analysis}, 45(2):159--178,
  2004.

\bibitem{tewes2013}
M.~{Tewes}, F.~{Courbin}, and G.~{Meylan}.
\newblock {COSMOGRAIL: the COSmological MOnitoring of GRAvItational Lenses. XI.
  Techniques for time delay measurement in presence of microlensing}.
\newblock {\em Astronomy \& Astrophysics}, 553:A120, May 2013.

\bibitem{bonvin2016}
V.~{Bonvin}, M.~{Tewes}, F.~{Courbin}, T.~{Kuntzer}, D.~{Sluse}, and
  G.~{Meylan}.
\newblock {COSMOGRAIL: the COSmological MOnitoring of GRAvItational Lenses. XV.
  Assessing the achievability and precision of time-delay measurements}.
\newblock {\em Astronomy \& Astrophysics}, 585:A88, January 2016.

\bibitem{howard2011}
A.~W. {Howard}, J.~A. {Johnson}, G.~W. {Marcy}, D.~A. {Fischer}, J.~T.
  {Wright}, G.~W. {Henry}, H.~{Isaacson}, J.~A. {Valenti}, J.~{Anderson}, and
  N.~E. {Piskunov}.
\newblock {The NASA-UC Eta-Earth Program. II. A Planet Orbiting HD 156668 with
  a Minimum Mass of Four Earth Masses}.
\newblock {\em \apj}, 726:73, January 2011.

\bibitem{bouchy2013}
F.~{Bouchy}, R.~F. {D{\'{\i}}az}, G.~{H{\'e}brard}, L.~{Arnold}, I.~{Boisse},
  X.~{Delfosse}, S.~{Perruchot}, and A.~{Santerne}.
\newblock {SOPHIE+: First results of an octagonal-section fiber for
  high-precision radial velocity measurements}.
\newblock {\em Astronomy \& Astrophysics}, 549:A49, January 2013.

\bibitem{bertaux2012}
J.~L. {Bertaux}, R.~{Lallement}, S.~{Ferron}, C.~{Boonne}, and R.~{Bodichon}.
\newblock {TAPAS, a web-based service of atmospheric transmission computation
  for astronomy}.
\newblock In {\em {ASA/HITRAN conference}}, 2012.

\bibitem{bertaux2014}
J.~L. {Bertaux}, R.~{Lallement}, S.~{Ferron}, C.~{Boonne}, and R.~{Bodichon}.
\newblock {TAPAS, a web-based service of atmospheric transmission computation
  for astronomy}.
\newblock {\em Astronomy \& Astrophysics}, 564:A46, April 2014.

\bibitem{bouchy2004}
F.~{Bouchy}, C.~{Lovis}, M.~{Mayor}, F.~{Pepe}, D.~{Queloz}, S.~{Udry},
  C.~{Melo}, and N.~{Santos}.
\newblock {New Results on Doppler Follow-up of Planetary Companions Detected by
  OGLE}.
\newblock In J.~{Beaulieu}, A.~{Lecavelier Des Etangs}, and C.~{Terquem},
  editors, {\em Extrasolar Planets: Today and Tomorrow}, volume 321 of {\em
  Astronomical Society of the Pacific Conference Series}, page~15, December
  2004.
  

  

\end{thebibliography}

\bibliographystyle{unsrt}

\end{document}